\documentclass[journal=dce]{CUP-JNL-DTM}%

\usepackage{graphicx}
\usepackage{multicol,multirow}
\usepackage{amsmath,amssymb,amsfonts}
\usepackage{mathrsfs}
\usepackage{amsthm}
\usepackage{rotating}
\usepackage{appendix}
\usepackage{ifpdf}
\usepackage[T1]{fontenc}
\usepackage{newtxtext}
\usepackage{newtxmath}
\usepackage{textcomp}
\usepackage{xcolor}
\usepackage[colorlinks,allcolors=blue]{hyperref}

\theoremstyle{definition}

\numberwithin{equation}{section}

\jname{Data-Centric Engineering}
\articletype{Research Article}
\jyear{2026}

\begin{document}

\begin{Frontmatter}

\title{Distribution-Free Conformal Prediction for Steel Fatigue Strength: Marginal Validity Is Not Enough}

\author[1]{Irene Boruah}

\authormark{Boruah}

\address[1]{\orgdiv{Department of Mechanical Engineering}, \orgname{Dibrugarh University Institute of Engineering and Technology}, \orgaddress{\city{Dibrugarh}, \postcode{786004}, \state{Assam},  \country{India}}. Corresponding author: ireneboruah2004@gmail.com}

\keywords{conformal prediction, uncertainty quantification, steel fatigue strength, conditional coverage, gradient boosting, materials informatics}

\abstract{Predicting fatigue failure in steel components experimentally is costly, requiring testing across multiple compositions and processing conditions, which has spurred research on data-driven prediction models. Studies using the NIMS MatNavi steel fatigue dataset often report high point-prediction accuracy but rely on aggregate error metrics, leaving uncertainty about the reliability of individual predictions and whether accuracy is consistent across the fatigue-strength spectrum. This paper is the first to apply conformal prediction to steel fatigue strength, comparing seven interval-construction methods across 50 independent data splits and distinguishing marginal coverage from coverage within specific sub-regions of the predicted property. A gradient-boosting point model achieves an $R^2$ of $0.976 \pm 0.009$ and a mean absolute error of $18.3 \pm 2.3$~MPa. Split-conformal prediction provides valid marginal coverage (0.918) but drops to 0.758 in the highest-strength quartile, where design margins are most critical, a pattern also observed with a Gaussian process baseline. Two locally-adaptive methods correct this: a cross-fitted normalized conformal method holds 0.872--0.940 across quartiles at no cost in average width, and Mondrian group-conditional conformal prediction holds the tightest band of any method (0.917--0.939) at a 12\% width premium, part of which traces to the more conservative finite-sample quantile level implied by per-group calibration at this sample size. Conformalized quantile regression, by contrast, restores marginal validity but inflates intervals in every quartile without closing the conditional gap. Marginal coverage claims for ML-based fatigue-strength predictions can conceal systematic unreliability precisely where engineering decisions are most risky; therefore, conditional coverage should be routinely assessed alongside marginal coverage.}

\end{Frontmatter}

\section*{Impact Statement}
Steel fatigue failure causes most structural and mechanical component breakdowns, and machine learning now predicts fatigue strength accurately from composition and processing data, avoiding months of physical testing. But accuracy is not trustworthiness: an engineer setting a design allowable needs a prediction interval that holds up where it matters, not just on average. This work shows standard uncertainty methods are unreliable in the high-strength regime, where margins are tightest and failure is costliest. We identify two methods that correct this, one at no added width and one at a modest width cost, explaining rather than hiding the uncertainty that remains. This offers engineers a concrete check against false confidence: verify coverage holds locally, not only on average.

\section{Introduction}\label{sec:intro}

Fatigue failure, cracking under repeated stress well below a material's ultimate strength, is one of the most common ways steel components fail in service, and one of the hardest to catch in advance, since there is typically no visible deformation warning before it happens. Determining fatigue strength experimentally means running physical tests over many load cycles, which is slow and expensive to repeat across composition and processing variants. That cost has pushed researchers toward data-driven prediction instead: train a model on existing fatigue data, then predict strength for a new composition and processing route without physical testing.

That approach works well. On the NIMS MatNavi steel fatigue dataset used throughout this paper, published models already reach $R^2$ above 0.95 \citep{agrawal2018,liu2023,kookalani2025}, and recent work has pushed into interpretability, using SHAP to identify which features drive predictions and, in Liu et al.'s case, symbolic regression to recover a closed-form expression. What none of this work does is say how much to trust any individual prediction: every study reports $R^2$ and mean absolute error, aggregate accuracy, and stops there. An engineer setting a safety margin needs more, not just what the model predicts, but how wide the real uncertainty band is, and whether that interval has earned its stated confidence.

Gaussian process regression is the usual answer, but its uncertainty estimate assumes a specific noise structure with no guarantee that assumption matches real fatigue data. Conformal prediction offers a distribution-free alternative: a 90\% interval that really does contain the true value 90\% of the time, without assuming anything about the noise distribution, so long as calibration and test data are exchangeable. But that guarantee is usually only checked as an average over the whole test set, marginal coverage. A method can hit 90\% on average while being badly wrong in specific sub-regions, and a marginal-only check would never reveal it. For fatigue strength, this matters more than usual: the sub-region most likely to fail, high-strength steel, is also where design margins are tightest.

This paper compares seven interval-construction methods for steel fatigue strength across 50 data splits, separating marginal coverage from coverage within sub-regions of the predicted property. Split-conformal prediction and a Gaussian process baseline are both marginally valid yet conditionally miscalibrated, badly under-covering the high-strength quartile where design margins matter most. Two locally-adaptive methods repair this by different mechanisms and at different costs: a cross-fitted normalized conformal method restores near-uniform coverage at essentially no width cost, and Mondrian group-conditional conformal prediction achieves the tightest conditional band of any method tested at a modest width premium. Conformalized quantile regression, which conformalizes an unreliable quantile baseline, restores marginal validity without restoring conditional calibration. We diagnose the mechanisms behind each of these outcomes rather than leaving them unexplained.

Section~\ref{sec:related} reviews prior work. Section~\ref{sec:methods} describes the dataset, model, and methods compared. Section~\ref{sec:results} presents results, marginal coverage first, then the conditional-coverage finding that drives the rest of the paper. Section~\ref{sec:conclusion} concludes.

\section{Related Work}\label{sec:related}

\subsection{Data-driven fatigue strength prediction}

\citet{agrawal2014} were the first to apply supervised learning to the NIMS MatNavi fatigue dataset, using composition and processing parameters to predict fatigue strength directly rather than relying on physics-based models that are slow and expensive to build. \citet{agrawal2018} turned this into a practical ensemble data-mining tool. \citet{liu2023} pushed accuracy on the same dataset substantially further, using engineered atomic-level features (mean electronegativity, valence-electron counts) alongside composition and processing, and identifying tempering temperature and mean electronegativity as the dominant SHAP predictors; they additionally introduce symbolic regression to recover a closed-form expression linking their four key features to fatigue strength. \citet{kookalani2025} work with the same 437-sample dataset and, closer to our own feature choices, raw composition rather than engineered atomic descriptors: their CatBoost model reaches $R^2 = 0.952$, and SHAP identifies tempering temperature, chromium, and molybdenum as dominant predictors. What is consistent across all four studies is the evaluation: $R^2$ and MAPE, full stop. Even the two that go furthest toward interpretability, Liu et al.\ with SHAP and symbolic regression, Kookalani et al.\ with SHAP alone, report point-prediction accuracy only: no study in this line constructs a prediction interval with any coverage guarantee, and none checks whether that accuracy holds up evenly across the fatigue-strength range or falls apart somewhere. That is the gap we are after.

\subsection{Uncertainty quantification in materials machine learning}

Outside the fatigue-specific literature, a wider materials-informatics community has been working on UQ for property prediction generally. Gaussian process regression is the default choice, mostly because it comes with a posterior variance built in, but it assumes a noise structure (usually homoscedastic, sometimes a simple heteroscedastic extension) that may not actually match the property it is applied to. \citet{tavazza2021} compared quantile-loss, direct error-learning, and GP approaches across several material properties. \citet{varivoda2023} ran a broader benchmark of UQ methods for materials property prediction. Varivoda et al.\ is the one that matters most here, because it includes inductive conformal prediction. As far as we have found, it is the only materials-property UQ benchmark that does. But it stops at marginal coverage and interval efficiency. Whether coverage holds up within sub-regions of the predicted property is not asked. That question is still open, and it is the one we take up.

\subsection{Conformal prediction and the marginal--conditional coverage distinction}

Conformal prediction \citep{vovk2005} gives prediction intervals a distribution-free, finite-sample marginal coverage guarantee, with the only real requirement being exchangeability between calibration and test data. No parametric noise assumption needed. Split-conformal prediction \citep{lei2018} makes this workable at scale: one held-out calibration set instead of the full conformal procedure, at the cost of a single, constant interval width applied everywhere. \citet{papadopoulos2008} introduced normalized nonconformity measures as a fix for exactly this: scaling the nonconformity score by a predicted difficulty estimate so interval width can vary locally instead of staying fixed, which is the technique our own method builds on directly (Section~\ref{sec:uq}). That constant width, before normalization, is exactly the weak point that motivated locally-adaptive variants generally. A separate line of work partitions the data into predefined groups and calibrates within each, so the coverage guarantee applies per group rather than only in aggregate. \citet{vovk2005} describe this Mondrian construction alongside the general conformal framework; it trades a smaller calibration set per group for a guarantee that holds within each group by construction, which makes it the natural point of comparison for any claim about conditional coverage. \citet{vovk2012} first laid out, rigorously, what ``conditional'' validity can and cannot mean for a conformal predictor, and showed that inductive conformal predictors are only known to control coverage on average, not within arbitrary sub-populations, which is precisely the gap normalization and later methods try to close. \citet{romano2019} built conformalized quantile regression (CQR) to address it from a different angle, pairing conformal calibration with quantile regression so interval width tracks local difficulty. They showed, both in theory and in practice, that a method can be marginally valid while still badly miscalibrated within specific sub-populations. That marginal-versus-conditional distinction is the core problem we are bringing into the fatigue-strength setting, though our locally-adaptive mechanism is a cross-fitted normalized conformal score rather than CQR (Section~\ref{sec:uq}). It is worth being upfront about the ceiling here: \citet{barber2021limits} prove that exact conditional coverage is impossible to guarantee distribution-free for continuous targets, without further assumptions on the underlying distribution. Normalization and CQR are both practical approximations to conditional validity, not violations of that impossibility result, and we return to this point in Section~\ref{sec:limitations} when discussing the residual coverage gap our own method does not fully close.

\subsection{Positioning}\label{sec:positioning}

As far as we can tell, nobody has applied distribution-free conformal prediction to steel fatigue strength before. And within the broader materials-UQ literature, nobody has checked whether a conformal method's coverage holds conditionally across sub-regions of a predicted property. We address both. It is worth being precise about what this adds beyond \citet{liu2023} and  \citet{kookalani2025}, the two strongest existing works on this exact dataset: their SHAP-based analyses explain why the model predicts what it predicts, feature attribution and, in Liu et al.'s case, a closed-form expression, but neither says anything about how much to trust any individual prediction. Interpretability and calibrated uncertainty are different questions, and the second one has not been asked on this dataset. We show that marginally-valid split-conformal intervals fall apart in the high-strength regime specifically, where design margins are tightest. A cross-fitted normalized conformal method fixes most, though not all, of that failure, and a Mondrian group-conditional construction closes it entirely at a measurable cost in width.

\section{Materials and Methods}\label{sec:methods}

\subsection{Dataset}\label{sec:dataset}

We use the NIMS MatNavi steel fatigue dataset: 437 samples, 25 features covering chemical composition (C, Si, Mn, P, S, Ni, Cr, Cu, Mo) and processing (normalizing, through-hardening, tempering, carburizing, plus upstream rolling and cooling parameters). The target is rotating-bending fatigue strength, spanning 225 to 1{,}190~MPa across the dataset.

Each of the 50 runs (Section~\ref{sec:protocol}) uses its own random split into training, calibration, and test sets. Test is 20\% of the data. Of what is left, 25\% goes to calibration (used only for conformal calibration, Section~\ref{sec:uq}), and the rest trains the model. We fit feature standardization on the training set only, then apply it to calibration and test, so no scaling statistics leak across the split. A row-index column present in the source spreadsheet (a serial number with no physical meaning) is dropped before any model fitting, so it is absent from the 25 features above and from every result reported below.

\subsection{Data Leakage Audit}\label{sec:leakage}

With only 437 samples, we wanted to rule out train/test contamination as an explanation for the accuracy before trusting it. Three checks: full-row duplicates (identical feature and target values), feature-only duplicates (same composition and processing but different fatigue strength, a sign of label noise or measurement duplication rather than real replicates), and near-duplicates, flagged by scaled-Euclidean distance below 5\% of the dataset's median nearest-neighbor distance (threshold 0.0160, against a median of 0.3195).

We found none of the three. Given how high the $R^2$ is for a dataset this small, it is worth stating plainly: the accuracy in Section~\ref{sec:pointperf} reflects real generalization, not leakage.

\subsection{Point Prediction Model}\label{sec:pointmodel}

We predict fatigue strength with a gradient boosting regressor \citep{friedman2001}. We tuned hyperparameters once, on the seed-42 fold, then froze them for all 50 runs. Re-tuning per split would itself be a form of leakage. The frozen values are n\_estimators~=~200, learning\_rate~=~0.05, max\_depth~=~4, min\_samples\_leaf~=~1, subsample~=~1.0, with random\_state~=~0. Note that the tuning fold falls inside the 0--49 evaluation range, so one of the 50 splits is not fully held out with respect to hyperparameter selection; with a single tuning pass spread across 50 splits the effect on the reported means is negligible.

We used SHAP \citep{lundberg2017} to check interpretability. Chromium content (Cr) and tempering temperature (TT) were the two dominant predictors by a clear margin, followed by quenching media temperature (QmT), carbon content (C), and normalizing temperature (NT). That ranking lines up with what is already known metallurgically about what drives fatigue strength: Cr and TT govern hardenability and the tempering response that sets the final microstructure, while QmT controls the cooling rate during hardening and therefore the resulting martensite fraction, so their dominance here is not a surprising result, it is a sanity check the model passes. (Section~\ref{sec:pointperf} presents the SHAP figures alongside the other point-prediction performance results.)

\subsection{Uncertainty Quantification Methods}\label{sec:uq}

We compare seven ways of building prediction intervals at $1-\alpha = 0.90$. $\mu(x)$ is the point prediction from the gradient boosting model.

\paragraph{Split-Conformal Prediction (SCP).} The nonconformity score on the calibration set is the absolute residual, $s_i = |y_i - \mu(x_i)|$. We take the $(1-\alpha)$-quantile of these scores at level $\lceil (1-\alpha)(n+1) \rceil / n$ to get a single number, $\hat{q}$, and every test interval becomes $\mu(x) \pm \hat{q}$. Same width, everywhere.

\paragraph{Normalized Conformal Prediction, cross-fitted.} This is our method, adapting normalized nonconformity measures \citep{papadopoulos2008} to a cross-fitted difficulty estimate, and the interesting part is how it gets around SCP's constant width. We train a difficulty model $\rho(x)$ to predict expected absolute residual size, so interval width can vary locally rather than stay fixed. Here is the detail that matters: we train $\rho(x)$ on out-of-fold residuals from 5-fold cross-fitting on the training set, following the same logic that motivates cross-fitting in double/debiased machine learning more generally \citep{chernozhukov2018}: a model's residuals on its own training data are optimistically small, so any downstream estimate built from those residuals inherits that bias unless the two roles stay separated. We fit each fold's model on the other four and generate residuals on the held-out fold, so $\rho(x)$ never sees residuals from a model that saw those same points during training. Skip cross-fitting and use in-sample residuals instead, and they come out optimistically small, since a model always fits its own training data too well, and the method quietly collapses back into behaving like SCP. Cross-fitting makes the method work at all; without it, it is just SCP wearing a different name. We floor $\rho(x)$ at $0.1 \times$ the mean training residual to avoid dividing by something close to zero. The normalized score is $s_i = |y_i - \mu(x_i)| / \rho(x_i)$; its $(1-\alpha)$-quantile $\hat{q}_n$ comes from the calibration set; the interval is $\mu(x) \pm \hat{q}_n \cdot \rho(x)$, width scaling with local predicted difficulty.

\paragraph{Mondrian (group-conditional) split conformal.} We partition the predicted-strength range into four groups using cut points fixed from training data, then run split conformal separately within each. The cut points are the 25th, 50th, and 75th percentiles of the out-of-fold predictions generated by the same 5-fold cross-fitting used for $\rho(x)$ above, so no test information enters the group definition. Each calibration point is assigned to a group by $\mu(x_i)$; within group $g$ we take $\hat{q}_g$ as the $\lceil (1-\alpha)(n_g+1) \rceil / n_g$ quantile of the absolute residuals in that group, and a test point in group $g$ receives the interval $\mu(x) \pm \hat{q}_g$. Coverage then holds within each group by construction rather than only on average.

The cost is calibration-set size. Splitting roughly 87 calibration points four ways leaves about 22 per group, and the finite-sample correction $\lceil (1-\alpha)(n_g+1) \rceil / n_g$ is more conservative at $n_g = 22$ (effective level 0.954) than at the pooled $n = 87$ (effective level 0.920). The method therefore targets a higher level than nominal, and we return to this in Section~\ref{sec:limitations} when interpreting its over-coverage.

\paragraph{Conformalized quantile regression (CQR).} Following \citet{romano2019}, we conformalize the two pinball-loss models from the raw quantile-regression baseline below rather than fitting new ones. The nonconformity score is $E_i = \max\{\hat{q}_{\alpha/2}(x_i) - y_i,\ y_i - \hat{q}_{1-\alpha/2}(x_i)\}$, its $\lceil (1-\alpha)(n+1) \rceil / n$ quantile on the calibration set gives a single correction $Q$, and the interval is $[\hat{q}_{\alpha/2}(x) - Q,\ \hat{q}_{1-\alpha/2}(x) + Q]$. The correction is additive and global: one value of $Q$ applies to every test point regardless of where it sits in the predicted range.

The remaining three methods are baselines rather than our main contribution.

A \textbf{bootstrap ensemble} of 100 gradient boosting models, each fit on a bootstrap resample of training data with the same frozen hyperparameters, gives an interval as the empirical $[\alpha/2,\ 1-\alpha/2]$ percentile range across those 100 predictions at each test point. This captures epistemic uncertainty in the conditional mean only; it has nothing to say about the aleatoric scatter that is there regardless of how much data you collect.

\textbf{Gaussian process regression (GPR)} uses a constant kernel multiplied by an RBF kernel plus an additive white-noise term, fit on standardized features with internally normalized targets. Kernel hyperparameters are set by marginal-likelihood maximization using scikit-learn's default optimizer. The 90\% interval is $\mu_{GPR}(x) \pm 1.645 \cdot \sigma_{GPR}(x)$: a standard normal posterior interval.

\textbf{Raw quantile regression} \citep{koenker1978} trains two gradient boosting models directly on pinball loss, one targeting $\alpha/2$ and one targeting $1-\alpha/2$, with the same frozen hyperparameters minus the loss function. This baseline skips conformal calibration entirely. We include it to show what happens without that step.

\subsection{Evaluation Protocol}\label{sec:protocol}

All seven methods run across 50 independent train/calibration/test splits (seeds 0--49), hyperparameters frozen throughout (Section~\ref{sec:pointmodel}). Whatever variability shows up comes from how the data happens to split, not from re-tuning.

For each seed and method we compute marginal coverage (fraction of test points whose true fatigue strength lands inside the interval) and mean prediction interval width (MPIW). We report both as mean $\pm$ standard deviation across the 50 seeds (Table~\ref{tab:marginal}).

For conditional calibration, we partition each seed's test set into quartiles by predicted fatigue strength and compute coverage within each quartile, averaged over 50 seeds (Table~\ref{tab:conditional}), together with mean interval width in each quartile (Table~\ref{tab:width}). Quartile cut points are fixed from the out-of-fold training predictions described in Section~\ref{sec:uq} rather than from each seed's own test-set percentiles. Deriving the cut points from training data keeps the evaluation groups independent of the predictions being evaluated, and it is required for the Mondrian comparison, since that method calibrates on training-derived groups and would otherwise be scored against a different partition than the one it calibrated on. This diagnostic is what catches a method being right on average while wrong in specific places.

Where Q4 (the highest-strength quartile) shows conditional miscalibration, we dig into why. We split the Q4 test residuals into a bias component and a dispersion component, then compare Q4's residual standard deviation against the pooled standard deviation in Q1--Q3.

\section{Results and Discussion}\label{sec:results}

\subsection{Point Prediction Performance}\label{sec:pointperf}

Across 50 independent train/test splits with frozen hyperparameters, the gradient boosting model achieved a coefficient of determination of $R^2 = 0.976 \pm 0.009$ and a mean absolute error of $18.3 \pm 2.3$~MPa on held-out test data, over a fatigue-strength range of 225 to 1{,}190~MPa. We report the multi-seed average rather than any single split to avoid presenting a favorable draw as a headline result. The modest standard deviation across seeds indicates that predictive accuracy is stable and not an artifact of a particular partition.

A leakage audit of the dataset (Section~\ref{sec:leakage}) supports the reliability of this accuracy: we found no full-row duplicates, no feature-only duplicates (identical composition with differing fatigue strength), and no near-duplicate pairs below a scaled-Euclidean threshold of 5\% of the median nearest-neighbor distance. The reported accuracy therefore reflects genuine generalization rather than train/test contamination, a relevant point given the high $R^2$ obtained on a dataset of only 437 samples.

SHAP analysis of the fitted model identified chromium content (Cr) and tempering temperature (TT) as the two dominant predictors by a clear margin, followed by quenching media temperature (QmT), carbon content (C), and normalizing temperature (NT) as a second tier. This ranking is consistent with established metallurgical understanding: Cr and TT govern hardenability and the tempering response that sets the final microstructure, and QmT governs the cooling rate during hardening, which directly sets the martensite fraction, so their dominance supports the physical plausibility of the learned relationships rather than being an artifact of the model.

\begin{figure}[h]
    \centering
    \includegraphics[width=0.9\textwidth]{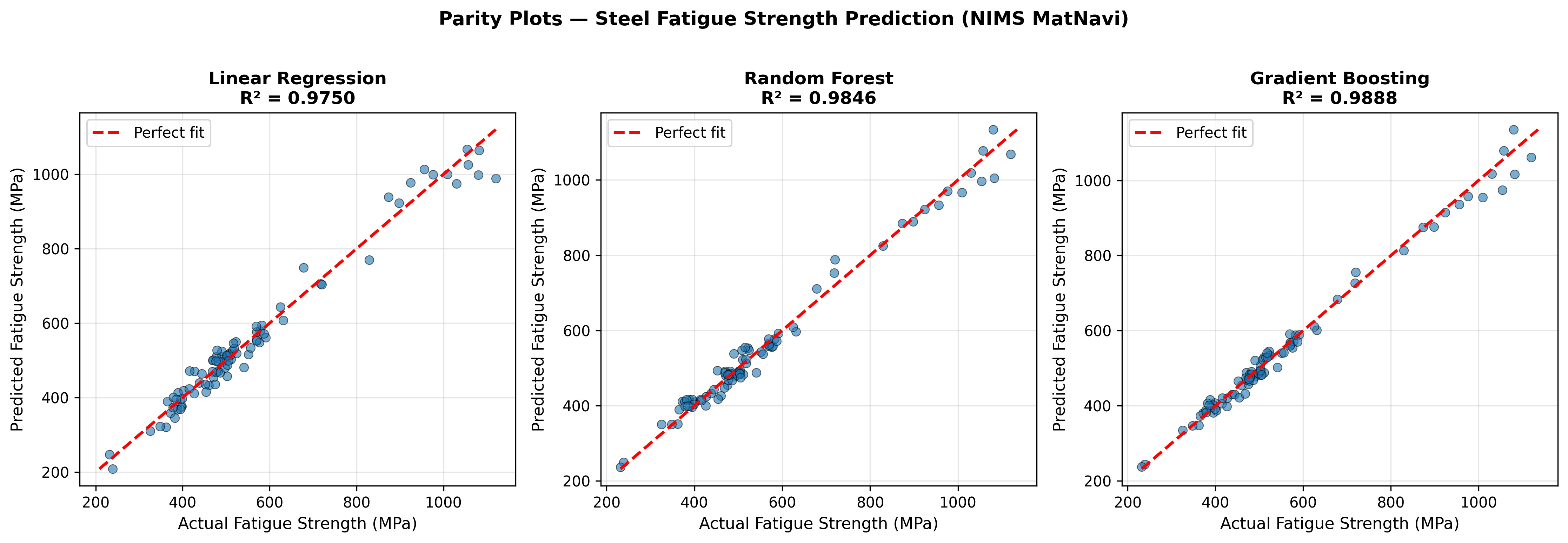}
    \caption{Parity plot, predicted vs.\ measured fatigue strength across three model types on a single representative split (seed 42), with the $y=x$ reference line. The headline accuracy reported above is the 50-seed mean and differs slightly from the single-split value shown here}
    \label{fig:parity}
\end{figure}

\begin{figure}[h]
    \centering
    \includegraphics[width=0.7\textwidth]{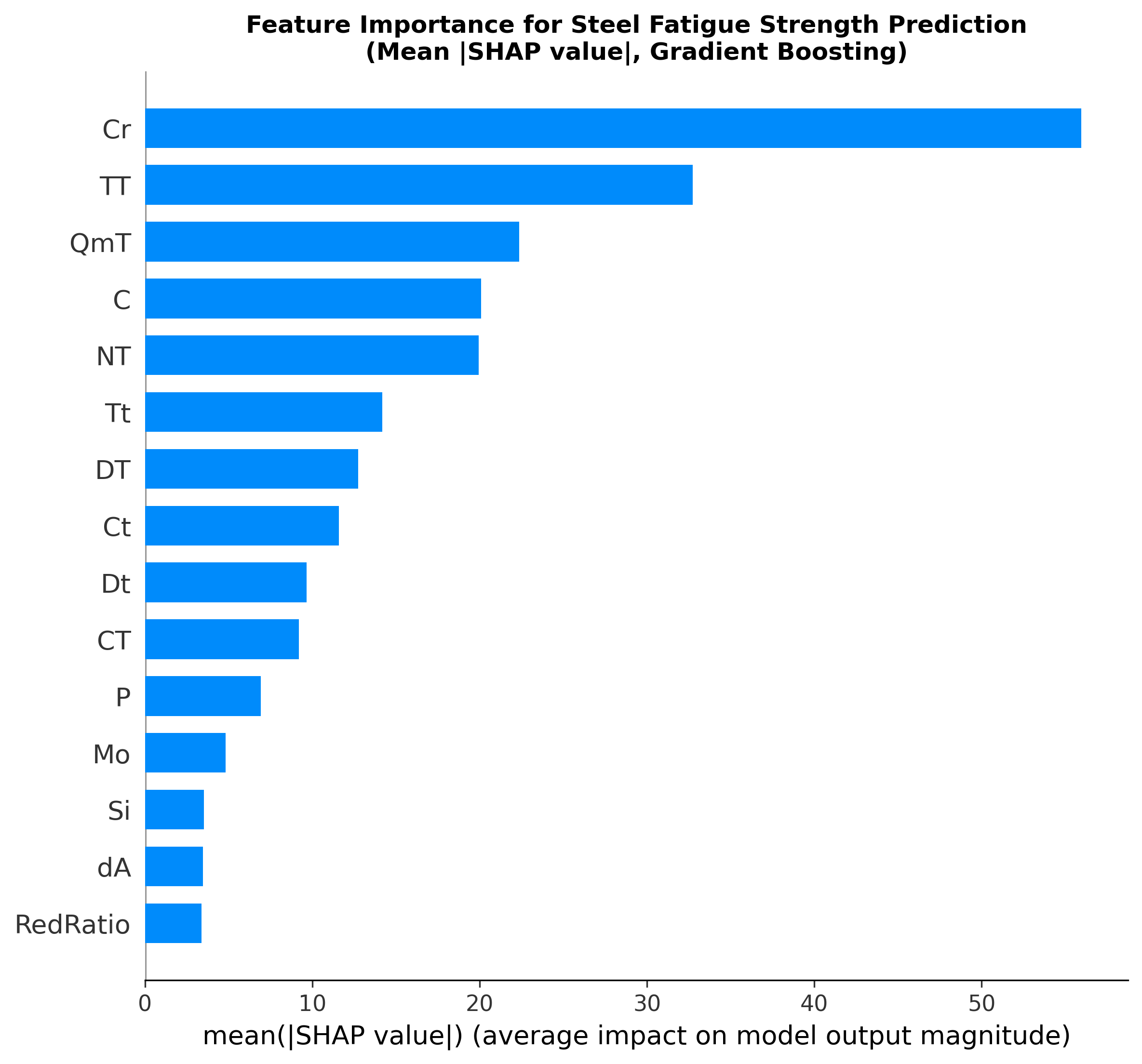}
    \caption{Mean absolute SHAP value per feature, ranked by importance}
    \label{fig:shap_bar}
\end{figure}

\begin{figure}[h]
    \centering
    \includegraphics[width=0.7\textwidth]{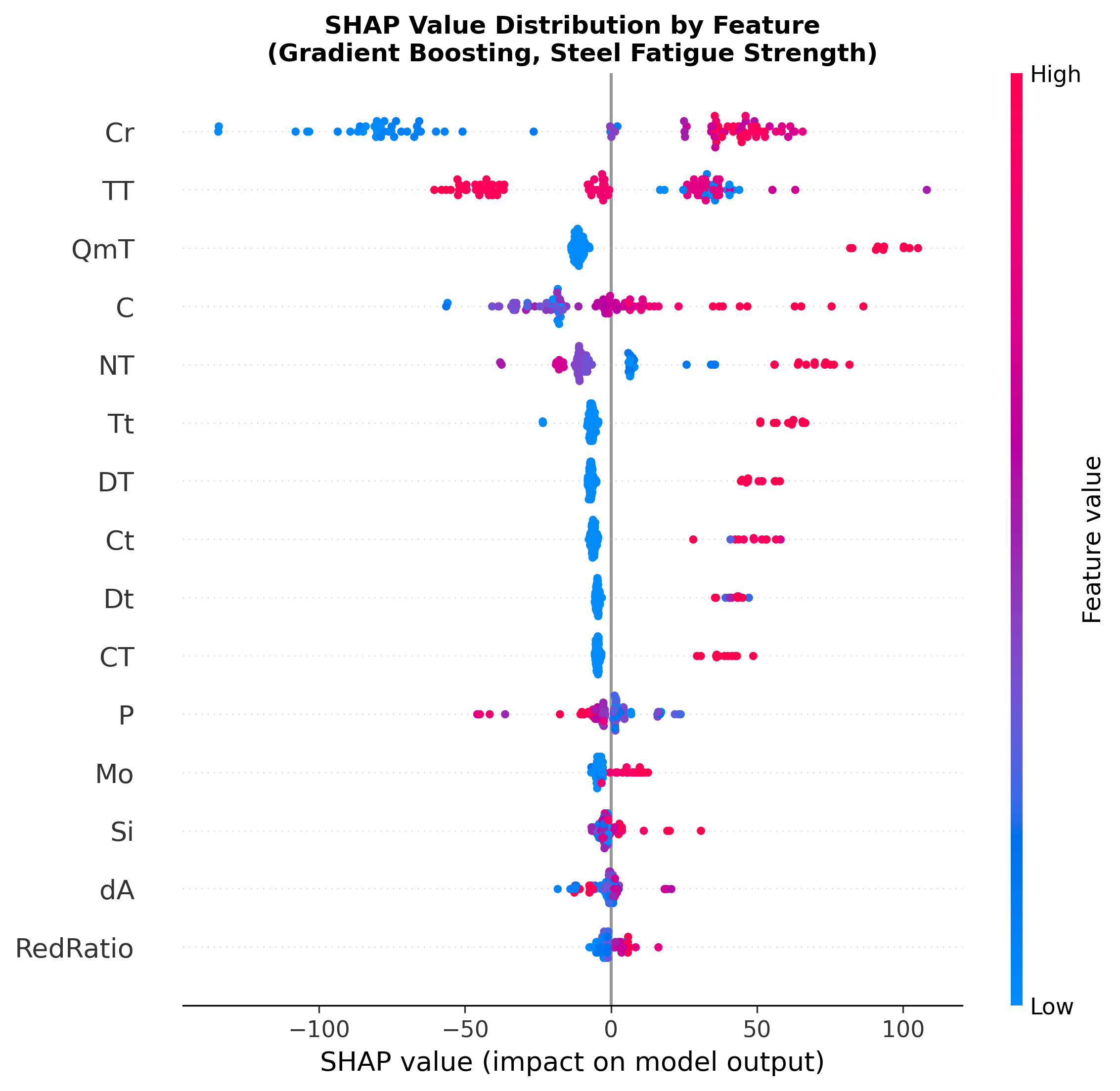}
    \caption{SHAP beeswarm plot showing the direction and spread of each feature's effect on predicted fatigue strength}
    \label{fig:shap_beeswarm}
\end{figure}

\subsection{Marginal Coverage of Prediction Intervals}\label{sec:marginalsec}

Deployment of fatigue-strength predictions in design requires not only accurate point estimates but reliable uncertainty quantification. We evaluated seven interval-construction methods at a target coverage of $1-\alpha = 0.90$, reporting empirical marginal coverage and mean prediction interval width (MPIW) averaged over the 50 seeds (Table~\ref{tab:marginal}).

\begin{table}[h]
\centering
\caption{Marginal coverage and interval width (50-seed mean $\pm$ std, target coverage 0.90)}
\label{tab:marginal}
\begin{tabular}{lcc}
\toprule
Method & Coverage & MPIW (MPa) \\
\midrule
Split-Conformal (SCP) & $0.918 \pm 0.039$ & $97.9 \pm 22.4$ \\
Normalized Conformal (cross-fitted) & $0.906 \pm 0.041$ & $95.0 \pm 16.9$ \\
Mondrian Conformal (group-conditional) & $0.930 \pm 0.029$ & $106.5 \pm 19.6$ \\
Conformalized Quantile Regression & $0.920 \pm 0.037$ & $200.3 \pm 25.3$ \\
Gaussian Process Regression & $0.887 \pm 0.043$ & $72.8 \pm 3.9$ \\
Bootstrap Ensemble & $0.650 \pm 0.063$ & $46.1 \pm 6.5$ \\
Quantile Regression (raw) & $0.765 \pm 0.058$ & $156.6 \pm 18.4$ \\
\bottomrule
\end{tabular}
\end{table}

Two of the methods fail to provide valid marginal coverage. The bootstrap ensemble covered only 0.650 of test samples: by quantifying epistemic uncertainty in the conditional mean while ignoring irreducible aleatoric scatter, it produces intervals that are far too narrow (46.1~MPa). Raw quantile regression undercovered at 0.765 despite producing the widest intervals (156.6~MPa), indicating that the unconformalized quantile estimates are themselves poorly calibrated at this sample size. Neither method is suitable for reliability-sensitive use.

The conformal methods achieved near-nominal marginal coverage (SCP 0.918, normalized 0.906), as expected from their distribution-free finite-sample guarantee under exchangeability \citep{vovk2005,lei2018}. The Gaussian process baseline \citep{rasmussen2006} fell modestly below target (0.887) with the narrowest intervals among the valid methods (72.8~MPa).

Conformalization repairs the raw quantile-regression baseline's marginal validity, raising coverage from 0.765 to 0.920, but the correction required to do so is substantial: the calibrated offset $Q$ averages $21.8 \pm 9.3$~MPa and is added to both interval endpoints, producing a mean width of 200.3~MPa. CQR is the widest of the marginally valid methods by a factor of two. Mondrian conformal prediction reaches 0.930 with a mean width of 106.5~MPa, over-covering relative to the 0.90 target for the finite-sample reason given in Section~\ref{sec:uq}. On marginal coverage alone, four methods are defensible and the ordering among them is uninformative; the differences that matter appear only under conditional analysis.

\subsection{Conditional Coverage Analysis}\label{sec:conditionalsec}

Marginal coverage averaged over the full test distribution can mask systematic miscalibration within sub-regions of the input space. We partitioned each test set into quartiles by predicted fatigue strength using training-derived cut points and computed coverage within each (Table~\ref{tab:conditional}), together with mean interval width in each quartile (Table~\ref{tab:width}). The bootstrap ensemble and raw quantile regression are omitted from this analysis: neither achieves valid marginal coverage (Table~\ref{tab:marginal}), so conditional coverage is not a meaningful question for them. The five remaining methods separate into three groups.

\begin{table}[h]
\centering
\caption{Coverage by predicted-fatigue-strength quartile (50-seed mean $\pm$ std, target 0.90). Quartile cut points fixed from training-set out-of-fold predictions}
\label{tab:conditional}
\begin{tabular}{lccccc}
\toprule
Quartile & SCP & Normalized CP & Mondrian & CQR & GPR \\
\midrule
Q1 (lowest)  & $0.979 \pm 0.032$ & $0.907 \pm 0.077$ & $0.928 \pm 0.076$ & $0.908 \pm 0.085$ & $0.974 \pm 0.038$ \\
Q2           & $0.974 \pm 0.033$ & $0.940 \pm 0.055$ & $0.939 \pm 0.063$ & $0.988 \pm 0.030$ & $0.942 \pm 0.060$ \\
Q3           & $0.966 \pm 0.040$ & $0.901 \pm 0.072$ & $0.917 \pm 0.078$ & $0.971 \pm 0.039$ & $0.898 \pm 0.063$ \\
Q4 (highest) & $0.758 \pm 0.109$ & $0.872 \pm 0.075$ & $0.935 \pm 0.066$ & $0.816 \pm 0.089$ & $0.739 \pm 0.098$ \\
\midrule
Range        & 0.221 & 0.068 & 0.022 & 0.172 & 0.235 \\
\bottomrule
\end{tabular}
\end{table}

\begin{table}[h]
\centering
\caption{Mean interval width by predicted-fatigue-strength quartile (MPa, 50-seed mean $\pm$ std). SCP's width is constant across quartiles by construction}
\label{tab:width}
\begin{tabular}{lccccc}
\toprule
Quartile & SCP & Normalized CP & Mondrian & CQR & GPR \\
\midrule
Q1 (lowest)  & $97.9 \pm 22.4$ & $72.2 \pm 16.1$ & $72.7 \pm 26.6$ & $200.3 \pm 37.4$ & $73.4 \pm 8.0$ \\
Q2           & $97.9 \pm 22.4$ & $64.6 \pm 13.2$ & $73.8 \pm 29.9$ & $137.6 \pm 22.5$ & $64.0 \pm 4.6$ \\
Q3           & $97.9 \pm 22.4$ & $69.6 \pm 11.8$ & $73.3 \pm 19.8$ & $116.3 \pm 20.6$ & $63.6 \pm 5.5$ \\
Q4 (highest) & $97.9 \pm 22.4$ & $169.5 \pm 40.3$ & $200.5 \pm 59.0$ & $341.9 \pm 45.0$ & $89.1 \pm 8.9$ \\
\bottomrule
\end{tabular}
\end{table}

\paragraph{Constant-width methods fail in the high-strength quartile.} Split-conformal, despite valid marginal coverage, spans 0.758 to 0.979 across quartiles, a range of 0.221. It over-covers Q1--Q3 and under-covers Q4 by more than fourteen percentage points. The mechanism is visible in Table~\ref{tab:width}: SCP's width is 97.9~MPa everywhere by construction, too wide for the low-strength regime and too narrow for the high-strength one. The Gaussian process baseline shows the same failure with a range of 0.235 (0.974 in Q1, 0.739 in Q4), and its Q4 behaviour is the more troubling of the two, since it produces the narrowest Q4 intervals of any method (89.1~MPa) while achieving the worst Q4 coverage. Narrow intervals are not evidence of a well-calibrated model. The GP posterior varies smoothly with the inputs but does not track the actual residual scale in the high-strength regime, so its intervals are confident and wrong in exactly the place where confidence is least warranted.

\paragraph{Locally-adaptive conformal methods correct it, by two mechanisms and at two costs.} Cross-fitted normalized conformal prediction holds 0.872--0.940, a range of 0.068, and widens its Q4 intervals to 169.5~MPa while tightening Q1--Q3 to roughly 65--72~MPa. Mean width is 95.0~MPa, below SCP's 97.9. The method redistributes width rather than adding it. Mondrian conformal prediction achieves the tightest conditional band of any method tested, 0.917--0.939, a range of 0.022, and is the only method whose Q4 coverage exceeds the 0.90 target (0.935). Its Q4 width is 200.5~MPa against normalized conformal's 169.5, an 18\% premium, and its mean width is 106.5~MPa against 95.0, a 12\% premium.

The two methods reach conditional calibration by different routes. Normalized conformal learns a continuous difficulty function and scales a single global quantile by it, which is efficient but leaves a residual shortfall in Q4 (0.872 against the 0.90 target). Mondrian discards the continuous structure and calibrates a separate quantile within each group, which removes the shortfall entirely but forgoes the pooling that made the normalized method cheap. Part of Mondrian's advantage is not adaptation at all: with roughly 22 calibration points per group, the finite-sample level $\lceil (1-\alpha)(n_g+1) \rceil / n_g$ evaluates to 0.954 rather than the 0.920 obtained from the pooled calibration set, so the method targets a higher nominal level than the others. This accounts for its over-coverage in every quartile and for part of the width premium. The uniformity of its coverage across quartiles is a real property of the construction; the level at which that uniformity sits is partly an artifact of the sample size.

\paragraph{Conformal calibration alone is not sufficient.} CQR restores marginal validity (0.920) but does not restore conditional calibration: it spans 0.816 in Q4 to 0.988 in Q2, a range of 0.172, better than SCP but worse than either locally adaptive conformal method. It is also the widest method in every quartile, including 341.9~MPa in Q4 and 200.3~MPa in Q1, where the other valid methods sit near 72~MPa. The reason is structural. The pinball-loss quantile estimates are poorly calibrated pointwise at this sample size, as the raw baseline's 0.765 coverage already indicates, and CQR corrects them with a single additive offset applied uniformly. That offset is driven up by the regions where the underlying quantile estimates are worst, then applied everywhere, inflating intervals in regions that never needed the correction. A global correction cannot repair a locally varying defect. This is a statement about CQR at $n = 437$ with these base learners rather than about CQR in general; with better-calibrated quantile estimates the offset would be smaller and the method more competitive.

\begin{figure}[h]
    \centering
    \includegraphics[width=0.7\textwidth]{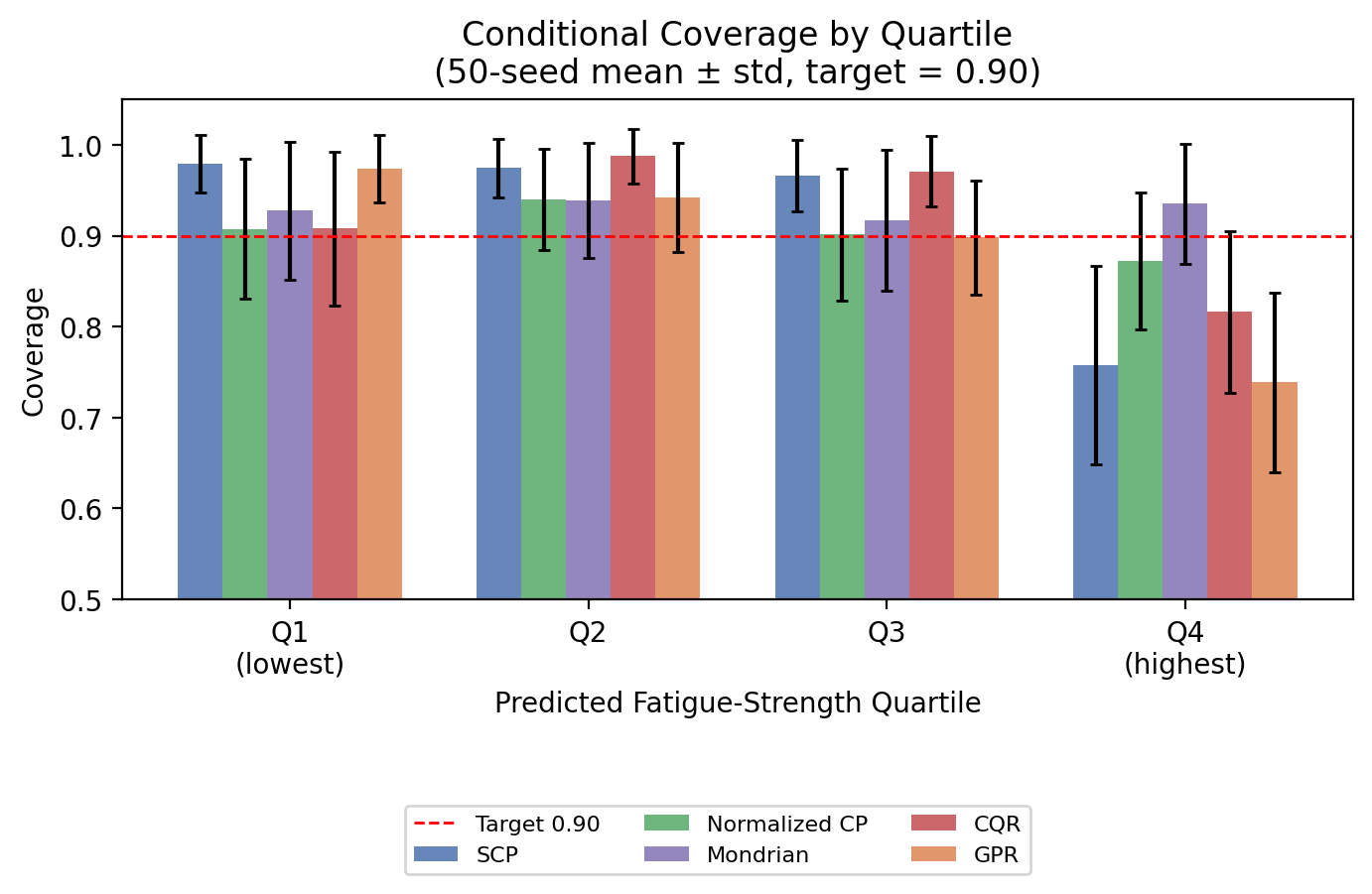}
    \caption{Grouped bar chart of coverage by quartile for SCP, normalized CP, Mondrian, CQR, and GPR, with the 0.90 target line}
    \label{fig:quartile_coverage}
\end{figure}

\begin{figure}[h]
    \centering
    \includegraphics[width=0.7\textwidth]{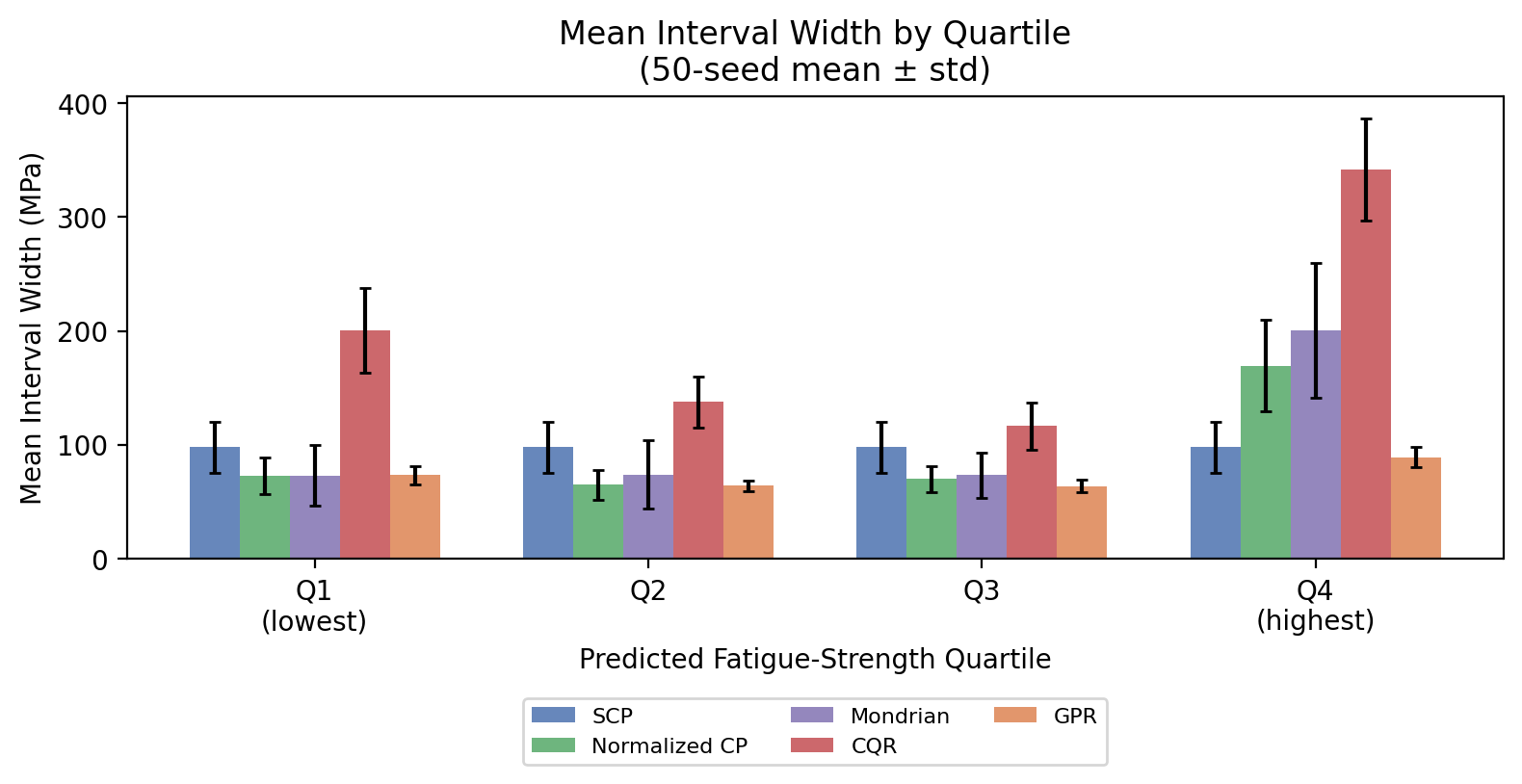}
    \caption{Grouped bar chart of mean interval width by quartile for SCP, normalized CP, Mondrian, CQR, and GPR}
    \label{fig:quartile_width}
\end{figure}

Taken together, these results establish that conditional coverage \citep{vovk2012,romano2019}, not marginal coverage alone, is the appropriate calibration criterion for this problem, and that satisfying it requires the interval width to vary locally. Of the seven methods compared, the two that vary width by a locally estimated quantity are the two that achieve it.

\subsection{Limitations}\label{sec:limitations}

Normalized conformal prediction does not fully close the coverage gap in the highest-strength quartile, where it reaches 0.872 against the 0.90 target. It is worth situating this against what is actually achievable: \citet{barber2021limits} prove that exact conditional coverage has no distribution-free guarantee for continuous targets without further distributional assumptions, so no method in this comparison could guarantee closing this gap. Mondrian's Q4 coverage of 0.935 is not a counterexample: it holds for the four groups as defined, not for arbitrary sub-populations, and comes at the higher effective level discussed below.

Diagnostic analysis of the Q4 residuals rules out a systematic directional bias as the driver: the mean signed residual there is essentially zero ($-0.1 \pm 9.4$~MPa across seeds), so the model is not consistently over- or under-predicting high-strength steels. The gap instead traces to elevated residual variance: Q4's residual standard deviation ($40.9 \pm 7.9$~MPa) runs roughly $2.7\times$ higher than the pooled Q1--Q3 standard deviation ($15.7 \pm 2.3$~MPa); computed per seed and averaged, the ratio is $2.68 \pm 0.70$. Cross-fitted normalization directly targets this heteroscedasticity by scaling interval width to a local difficulty estimate, which is why it closes most of the gap; the residual shortfall likely reflects the difficulty model's own estimation noise in a region where both training and calibration data are comparatively sparse, rather than a structural bias the method fails to account for.

Two limitations attach to the Mondrian comparison specifically. Its per-group calibration sets contain roughly 22 points, which makes the finite-sample quantile level conservative (0.954 against the pooled 0.920) and accounts for the method's uniform over-coverage; a larger calibration set would reduce both the over-coverage and the associated width premium, but we cannot verify by how much on a 437-sample dataset. Across the 50 seeds, 15 of 200 group-seed combinations contained fewer than 15 calibration points, all of them in Q1--Q3 and none in Q4, so the reported Q4 result does not depend on the conservative fallback applied in those cases. The group boundaries are also a modelling choice: quartiles are a natural partition for this diagnostic but not a principled one, and the guarantee holds for the groups as defined rather than for arbitrary sub-populations.

Remaining directions include localized conformal prediction with a kernel weighting over the calibration set (a direction suggested by Dr.\ Rohan Hore, personal communication), quantile base learners calibrated well enough for CQR to be competitive at this sample size, and enrichment of the dataset in the high-strength regime where samples are sparse.

\section{Conclusion}\label{sec:conclusion}

\begin{enumerate}
    \item Split-conformal prediction achieves nominal marginal coverage on steel fatigue strength (0.918) but is conditionally miscalibrated, ranging from 0.979 in the lowest-strength quartile to 0.758 in the highest.
    \item A Gaussian process baseline exhibits the identical failure pattern ($0.974 \rightarrow 0.739$), showing the problem is a property of the data's heteroscedastic structure, not an artifact of conformal construction specifically. It produces the narrowest high-strength intervals of any method while achieving the worst high-strength coverage.
    \item Two locally-adaptive conformal methods restore conditional calibration by different mechanisms. A cross-fitted normalized method holds 0.872--0.940 across quartiles at a mean width of 95.0~MPa, below the 97.9~MPa of split-conformal, by scaling a global quantile with a learned difficulty estimate. Mondrian group-conditional conformal prediction holds 0.917--0.939, the tightest band of any method tested, at a 12\% width premium, by calibrating a separate quantile within each group.
    \item Mondrian's uniformity is a property of its construction, but the level at which it sits reflects the sample size: with roughly 22 calibration points per group, the finite-sample quantile level is 0.954 rather than the 0.920 obtained from the pooled calibration set, which explains its over-coverage and part of its width cost.
    \item Conformal calibration alone does not deliver conditional coverage. Conformalized quantile regression achieves valid marginal coverage (0.920) but reaches only 0.816 in the highest-strength quartile while producing the widest intervals of any method in every quartile, because a single additive correction cannot repair quantile estimates that are miscalibrated by differing amounts in different regions.
    \item The residual coverage gap left by normalized conformal prediction in the highest-strength quartile (0.872 vs.\ the 0.90 target) traces to elevated residual variance in that regime, roughly $2.7\times$ the pooled Q1--Q3 level (mean per-seed ratio $2.68 \pm 0.70$), with no meaningful directional bias, rather than standing as an unexplained shortfall. This also reflects a proven theoretical limit \citep{barber2021limits}: exact conditional coverage has no distribution-free guarantee for continuous targets.
    \item Engineers using ML-based fatigue-strength predictions should treat marginal coverage claims with caution in the high-strength regime specifically, where design margins matter most and where this analysis shows standard methods are least reliable.
\end{enumerate}

\section*{Declarations}

\paragraph{Data availability statement.} The NIMS MatNavi steel fatigue dataset is available from Japan's National Institute for Materials Science (NIMS) subject to their access terms. The code used for data processing, model training, and analysis in this paper is openly available on Zenodo at \url{https://doi.org/10.5281/zenodo.21731736}.

\paragraph{Acknowledgments.} The author thanks Dr.\ Rohan Hore (Carnegie Mellon University) for a helpful discussion on region-specific and localized conformal prediction approaches to the conditional coverage gap discussed in Section~\ref{sec:limitations}.

\paragraph{Author contribution.} The author conceived the study, curated and audited the dataset, developed the methodology (including the cross-fitted normalized conformal method), performed the computational analysis, and wrote the manuscript.

\paragraph{Funding statement.} This research received no specific grant from any funding agency.

\paragraph{Competing interest.} None.

\paragraph{Ethical standard.} The research meets all ethical guidelines, including adherence to the legal requirements of the study country.

\printbibliography

\end{document}